\pdfoutput=1

\documentclass{vgtc}                          

\graphicspath{{figures/}{pictures/}{images/}{./}} 

\usepackage{times}                     
\usepackage{booktabs}                  
\usepackage{amsmath}                   
\usepackage{mathptmx}                  

\onlineid{0}
\vgtccategory{Research}
\vgtcinsertpkg

\title{The Choreographic Genome:\\Amplifying the Silent Structure of Text into Dance}

\author{Michael Li\thanks{e-mail: ml7@andrew.cmu.edu}
\and Alison Ding\thanks{e-mail: alisondi@andrew.cmu.edu}}
\affiliation{\scriptsize Carnegie Mellon University}

\teaser{
  \centering
  \includegraphics[width=\linewidth]{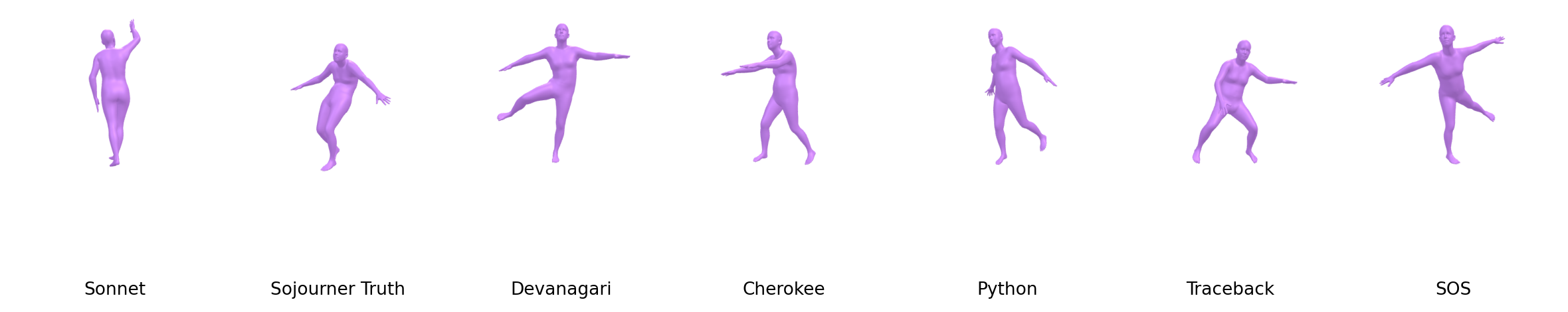}
  \caption{Seven texts, seven bodies. A single expressive keyframe from the
  choreography our instrument generates for each case-study text, from a
  Shakespeare sonnet to an ``SOS'' in Morse. The mapping from text to movement
  never consults meaning, only structure, yet each text still produces a visibly
  distinct dance. Every body is rendered from the same underlying engine.}
  \label{fig:teaser}
}

\abstract{
Recent advances in generative artificial intelligence have enabled the synthesis
of complex human motion with unprecedented fidelity. However, current
text-to-motion systems rely strictly on linguistic semantics: if an input reads
``I put my hands up'', the model searches for a pose with raised hands, and every
non-semantic property of the text is discarded as noise. In this work, we treat
that discarded structure as the signal. We present an embodied visualization
instrument that amplifies not what a text means, but how it is built. Our method
first quantizes dance kinematics into a motion codebook of 256 stylistic
``regions'' using Principal Component Analysis and K-Means clustering, and orders
those regions along the dominant axis of movement. We then map
the raw byte representation of any input text directly onto this codebook,
producing a deterministic sequence of regions that we call the text's
``choreographic genome''. A precomputed plausibility graph and a set of physics
smoothing routines turn this genome into fluid, full-body movement, so that the
dancing body becomes a display surface for the byte-level structure that semantic
systems ignore. Through a series of artistic case studies, including a Shakespeare
sonnet, a machine error log, source code, an abolitionist's question, and
Indigenous and Devanagari scripts, we show that each text produces a visibly
distinct dance, and that scripts marginalized by ASCII-centric computing are
amplified into close to three times as much movement per character. We frame this not
as a motion-synthesis benchmark, but as a critical and poetic visualization that
asks what we choose to count as signal, and what we allow to go unheard.
}


\begin{document}

\firstsection{Introduction}

\maketitle

While data permeates every aspect of our world, its representation is rarely equitable. Certain narratives are elevated to positions of influence, whereas others are left obscured, fractured, or intentionally suppressed~\cite{dignazio2020datafeminism,gitelman2013rawdata}. If visualization serves as a tool for amplification, bringing clarity to the quiet and presence to the ignored, we must critically examine how we wield it. In this paper, we explore this dynamic literally by interrogating our own creative systems: when we design a tool to magnify a specific signal, what exactly are we elevating, and what inadvertently gets muted?

Text is one of the most quietly structured data streams we produce. Every message
carries far more than its dictionary meaning; it carries an encoding, a
distribution of bytes, a choice of script, and a rhythm of repetition. However,
the dominant paradigm of generative artificial intelligence is built to hear only
one of these layers. Text-to-motion and text-to-image models treat the semantic
content of language as literal instruction, and they discard everything else as
noise~\cite{li2021aist,stanford2023edge}. For example, for typical choreography, if the lyrics say ``I put my hands up'', the model returns a figure with raised
hands. The structure of the language, which is the
actual material of it, is silenced in favor of its most sanctioned reading.

In this work, we present an embodied visualization instrument that inverts this
hierarchy. Rather than amplifying what a text means, we amplify how a text is
built. Our system reads the raw bytes of an input string and maps each byte
directly onto one of 256 quantized ``motion regions'', a learned vocabulary of
human movement. The resulting sequence is a deterministic structural fingerprint
that we call the text's ``choreographic genome'' (\cref{fig:genome}). A
precomputed plausibility graph then stitches the genome into biomechanically
continuous motion, and the dancing body becomes a display, a physical readout of
the byte-level structure that semantics-first systems throw away. Because the
mapping never consults meaning, every text, whether a sonnet, an error log, a
protest, a line of code, or a script that ASCII never planned for, is given a body
on equal terms (\cref{fig:teaser}).

We make three contributions. First, we reframe text-to-motion not as semantic
translation but as structural amplification, and we argue for the body as a
display surface for the otherwise invisible material of language
(\cref{sec:related,sec:system}). Second, we describe a fully functional and
interpretable pipeline, built on classical machine learning and graph search
rather than a black-box neural generator, that runs on a consumer CPU
(\cref{sec:system}). Third, through a set of artistic case studies over a
deliberately heterogeneous corpus, we show that the instrument gives each text a
visibly distinct dance, and that scripts marginalized by digital infrastructure
are amplified into close to three times as much movement per character as ASCII
English (\cref{sec:cases}).

We do not position this work as a more accurate motion model; the merit of our engine should not be judged based on typical motion generation quality metrics such as foot-skating ratio or floor penetration rate. We instead position this work as a
critical and poetic visualization that uses generative technology to amplify the
quiet structure of our data streams, and that keeps an honest account of the
biases it introduces along the way.

\section{Related Work}
\label{sec:related}

\subsection{Dance generation and the semantic ceiling}
Current research in AI dance generation focuses primarily on the technical
challenge of synthesizing high-fidelity kinetic movement, often conditioned on
text or audio. Systems such as EDGE, Bailando, and DanceFormer have achieved
remarkable success in generating physically plausible motion that adheres to a
provided prompt~\cite{stanford2023edge,siyao2022bailando,li2023danceformer},
trained on datasets such as AIST++, GDANCE, and
FineDance~\cite{li2021aist,le2023gdance,li2023finedance}. However, these systems
inherently bind movement to the literal, semantic meaning of their inputs, and
they treat dance generation as a supervised translation of human intent into
physical space. This approach inherits the cultural and linguistic biases of the
training data~\cite{crawford2021atlas}, and it leaves a critical gap regarding how
a generative system might interpret text abstractly. Our instrument targets
exactly this gap: it treats text as a structural sequence of raw choreographic
tokens rather than as a set of instructions to be obeyed.

\subsection{Amplification and what counts as signal}
The premise that ``raw data'' is ever truly raw, rather than shaped by those who
collect and encode it, has been thoroughly
critiqued~\cite{gitelman2013rawdata,dignazio2020datafeminism}. Drucker argues that
humanistic visualization should treat data as interpreted ``capta'' rather than
given fact~\cite{drucker2011humanities}, and Offenhuber's work on autographic
design attends to the material traces that data inscribes on the
world~\cite{offenhuber2023autographic}. Our work builds directly on this lineage.
The byte stream of a text is itself an indexical trace of an encoding standard, a
language, a keyboard, and an author, and amplifying that stream makes the trace
dance. This also connects our system to the post-digital ``aesthetics of failure''
and glitch traditions, which locate expressive signal in the noise and malfunction
that clean systems attempt to
suppress~\cite{cascone2000aesthetics,menkman2011glitch,church2017musicglitch,wallace2024breakingfromreality}.
Where a semantic model would discard this structure as merely noise, we treat it as the score.

\subsection{The body as display and the machine as co-creator}
Reading data off a moving body situates this work within data physicalization,
which studies how physical and bodily variables can carry information and support
reflection~\cite{jansen2015physicalization}. It also participates in a broader
shift in which AI moves from being a passive tool toward being a ``new neighbor'',
a social collaborator in curating collective memory and shared identity. For
example, projects such as Stephanie Dinkins' \emph{Not The Only One} use deep
learning to build oral archives, transforming the machine into a repository for
lived experience~\cite{dinkins2018ntoo}. In parallel, creative physical systems
such as the robot painter FRIDA demonstrate how an agent can bridge the digital
and physical worlds~\cite{schaldenbrand2022frida}. Like creativity research that
values deviation from learned norms over faithful
imitation~\cite{boden2004creative,elgammal2017can,jiang2024llmcreativity}, we seek
a co-creator rather than a mimetic parrot. Furthermore, because our method relies
on classical, interpretable algorithms rather than an opaque network, it remains
legible to the artists who use it, which makes it a far more suitable partner for
genuine creativity related collaboration.

\section{From Bytes to Bodies}
\label{sec:system}

Our framework operates through a multi-stage pipeline that shifts generation from
the predictive neural paradigm popular nowadays to a more classical, interpretable pathfinding approach. The system
has two phases: an offline phase that quantizes continuous motion into a
structural codebook and learns the physical rules for moving between its entries,
and a runtime phase that maps an input text directly onto that codebook and
synthesizes the corresponding choreography. Our full source code is publicly
available.

\subsection{The movement vocabulary}
We build the movement vocabulary from the AIST++ motion-capture
corpus~\cite{li2021aist}. First, we align each pose to a local, translation-invariant and
heading-invariant frame, so that global placement does not skew kinematic
comparisons. We then describe every frame by its local joint configuration, its
root and joint velocities, and a set of binary foot-contact labels, which together
yield 278 features per frame. Because we want each frame to be characterized by
the motion it belongs to rather than by a single static posture, we window these
descriptions over 20 consecutive frames, producing a 5560-dimensional motion
signature. In other words, we characterize each frame not only by the posture the body
holds at that instant, but also by the movement it performs over the next 20 frames.
We reduce this signature to 64 dimensions using
Principal Component Analysis (PCA), which retains the dimensions along which these frames
differ most, and we quantize the result into 256 clusters using K-Means, which
groups mutually similar frames together, giving exactly one cluster for each
possible byte value. Each cluster, which we call a
``region'', is a recurring stylistic gesture, and together the 256 regions form
the codebook, the instrument's alphabet of the body.

\subsection{Ordering the alphabet}
\label{sec:order}
K-Means returns clusters in an arbitrary order. If we mapped byte values directly
to these raw labels, the contiguous structures inherent to text encodings,
such as ASCII blocks or UTF-8 continuation bytes, would be obscured by random
kinematic leaps. Because our instrument aims to physicalize text structure, proximity
in the encoding should ideally translate to proximity in the body.

We therefore order the 256 codebook regions along the first principal component (PC1)
of the motion feature space, which accounts for $21.6\%$ of the variance. By definition
of the PCA algorithm, PC1 is the axis along which the regions differ most, and can hence
be interpreted as the dominant axis of our movement vocabulary.
As \Cref{tab:order} shows, this increases the rank correlation ($\rho$) between byte
distance and kinematic distance from $-0.01$ to $+0.41$ and reduces the mean kinematic
step between consecutive bytes by 22\%. We favor the use of PC1 over spectral layouts or
greedy nearest-neighbor walks because it establishes a global and interpretable order.
Consequently, texts drawn from narrow byte bands naturally constrain the body to
correspondingly narrow kinetic bands.

\begin{table}[tb]
\caption{Three byte-to-region indexing methods compared to the arbitrary baseline. $\rho$ is the rank correlation between byte distance and kinematic distance. Step is the mean kinematic distance between consecutive byte values. Our instrument uses PC1 rank.}
\label{tab:order}
\centering
\footnotesize
\begin{tabular}{@{}lrr@{}}
\toprule
Byte to region indexing & $\rho$ & Step \\
\midrule
Arbitrary K-Means labels     & $-0.01$ & 63.1 \\
\textbf{PC1 rank (ours)}     & $+0.41$ & 49.0 \\
Spectral (Fiedler) layout    & $+0.40$ & 50.5 \\
Greedy nearest-neighbor walk & $+0.32$ & 33.9 \\
\bottomrule
\end{tabular}
\end{table}

\begin{figure*}[!t]
  \centering
  \includegraphics[width=0.95\linewidth]{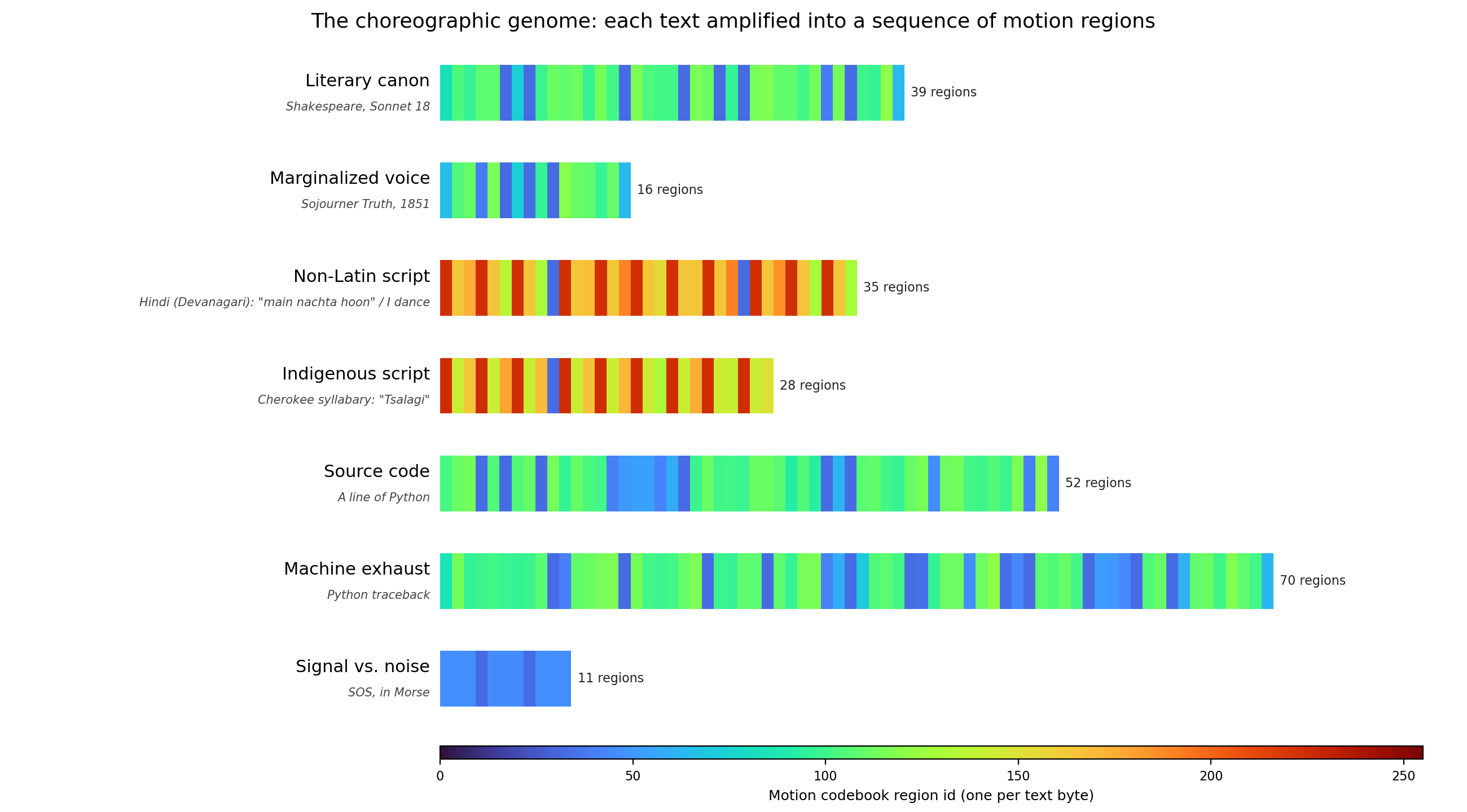}
  \caption{The ``choreographic genome''. Each text is amplified, byte by byte, into
  a sequence of motion regions, where color encodes region ID (\cref{eq:amp}).
  Because UTF-8 spends more bytes on non-Latin scripts, the Devanagari and Cherokee
  strips run long and warm despite their short character counts, while the ``SOS''
  distress call collapses into a brief, repetitive pulse. Each text yields a
  visibly distinct genome, with a mean pairwise edit distance of 44.3 across the
  corpus.}
  \label{fig:genome}
\end{figure*}

\subsection{The amplification operator}
The core of the system is intentionally, almost provocatively, simple. Let a text
be its UTF-8 byte sequence $b_1 b_2 \cdots b_n$, where each $b_i$ lies in
$[0, 255]$. The choreographic genome is the identity map from bytes to regions:
\begin{equation}
\Phi(b_1 b_2 \cdots b_n) = r_1 r_2 \cdots r_n, \qquad r_i = b_i .
\label{eq:amp}
\end{equation}
There is no learning in this step, no inference of intent, and no semantic lookup,
and that is precisely the point: the map amplifies structure because it refuses to
interpret. Two texts that mean the same thing but are built differently produce
different genomes. Furthermore, a text written in a multibyte script produces a
longer genome than its character count would suggest, because the encoding itself
is part of the signal we have chosen to hear.

One consequence of \cref{eq:amp} should be acknowledged. The
vocabulary contains 256 entries because a byte has 256 values, and not because 256
is the number that best partitions human movement. Sweeping $K$ from 16 to 1024,
we observe that the residual quantization error, which measures how much of the movement the
representative motions fail to capture, falls from $56\%$ of the variance
to $18\%$ very smoothly, while cluster separation, measured by silhouette, never
exceeds $0.16$ at any resolution we tested. The motion space therefore offers no
natural number of regions to be discovered, and our choice of 256 regions is best described as being externally motivated. The encoding, and not the motion data, sets the resolution of the body's alphabet.

\subsection{Making the genome danceable}
Adjacent bytes in a text routinely name gestures that no human could perform back
to back. Rather than smoothing this discontinuity away invisibly, we treat
physical continuity as an explicit and auditable layer. Offline, we precompute a
directed ``plausibility graph'' over the 256 regions. We define the cost of a
transition using a motion-matching heuristic that compares the pose, the joint and
root velocities, and the estimated root trajectory 15 and 30 frames into the
future, with an additional penalty on root velocity to prevent abrupt momentum
changes. We store a directed edge between two regions when this cost falls below a
fixed tolerance for a sufficient fraction of sampled frame pairs, and we weight the
edge by the average transition cost. Essentially, two regions are neighbors
if transitioning between them maintains physical continuity, meaning limb
positions, speeds, and trajectories align smoothly.
We keep this edge if the transition is plausible across most
sampled frames, recording its average cost. At runtime, when the genome demands a
transition that the body cannot make directly, we apply Dijkstra's algorithm over
this graph to insert the shortest sequence of intermediate ``bridge'' regions, and
we fall back to the nearest reachable region only when no path exists. Finally, we
ease the disjoint segments together: positional values are interpolated with a
Perlin smootherstep polynomial, joint rotations are blended with Spherical Linear
Interpolation to avoid limb collapse, and a Savitzky-Golay filter together with an
anti-skating correction removes residual jitter and foot sliding. Together, these
steps create smooth transitions, prevent limbs from intersecting the
body by rotating them along the shortest arc, and lock the feet to stop
the dancer from sliding. The result
honors the inherent structure of the input text while remaining biomechanically fluid. Crucially,
every step is inspectable: the genome, the injected bridges, and the physics
corrections are all logged, so the instrument can explain itself rather than
asking to be blindly trusted.

\begin{figure*}[!t]
  \centering
  \includegraphics[width=0.92\linewidth]{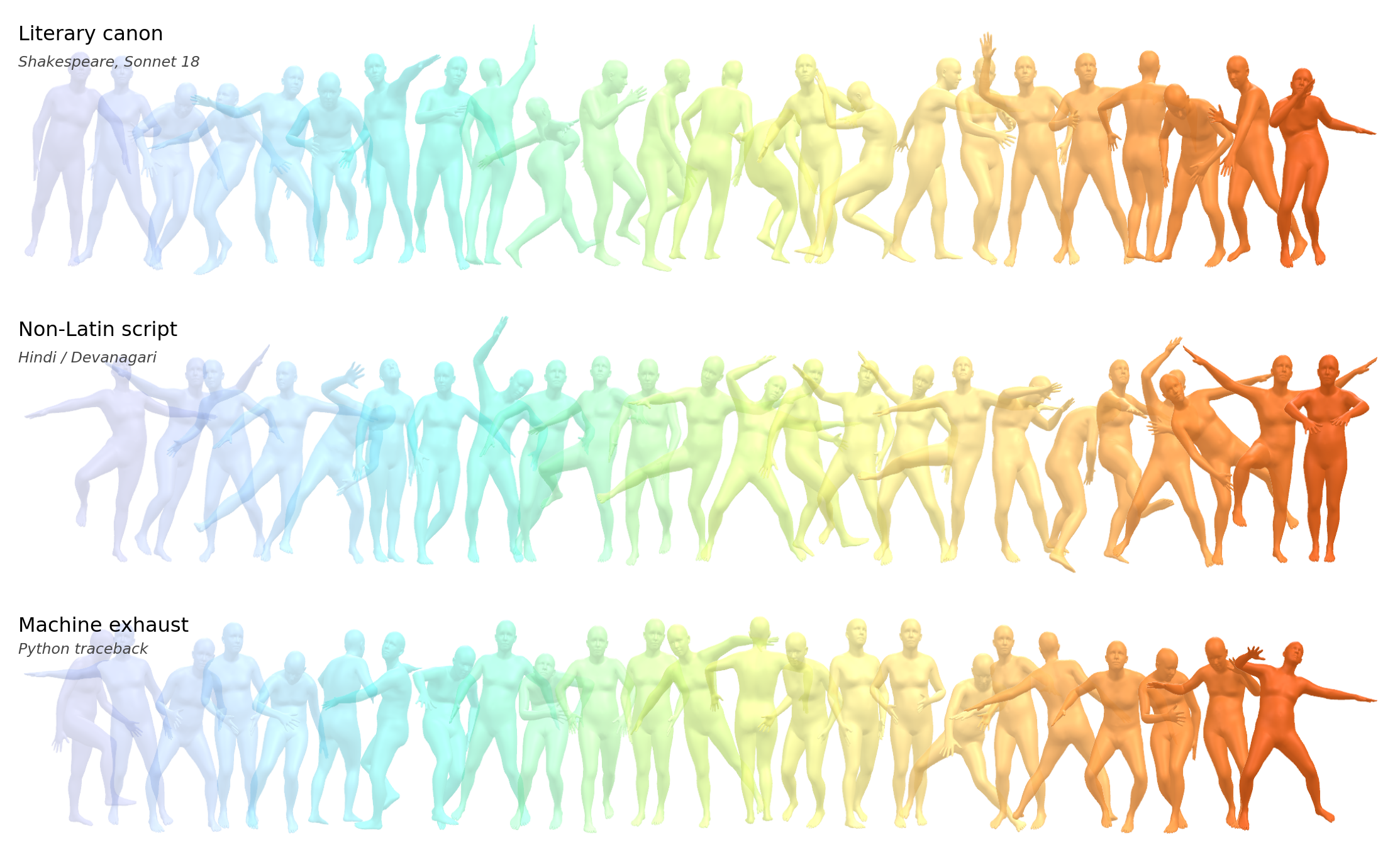}
  \caption{A chronophotographic view of three generated dances. We sample each
  sequence across time, spread the frames from left to right, and color them from
  cool to warm by moment. Each sweep is produced from byte structure alone: a
  Shakespeare sonnet (top), the Devanagari phrase meaning ``I dance'' (middle), and
  a Python traceback (bottom). The instrument never reads those meanings, yet the
  structure alone yields three visibly different choreographic arcs. Every sweep is
  sampled to the same number of frames, so the figure compares the character of the
  three dances and not their length.}
  \label{fig:trail}
\end{figure*}

\section{Amplifying Texts: Artistic Case Studies}
\label{sec:cases}

To study the instrument as a visualization tool, we
assembled a deliberately heterogeneous corpus of seven texts. Each entry comes
from different domains, testing our claim that structure alone can give a text
a body. Two are plain ASCII English prose at opposite ends of the canonical
spectrum: a line of Shakespeare and Sojourner Truth's ``Ain't I a woman?''.
Semantically they are worlds apart, but structurally they should be treated
similarly. Two use scripts that UTF-8 encodes with multiple bytes per
glyph, Devanagari and Cherokee, isolating the contribution of the encoding
standard itself. Two are machine registers, a Python line and a
traceback, testing whether text never meant to be read aloud still yields a
distinct body. Finally, ``SOS'' in Morse carries almost no lexical content,
asking what remains to be amplified when there is originally little to interpret.

For each text, we compute its genome and measure the movement required to
physically realize it. \Cref{tab:cases} reports these results across the four
categories, visualized in \cref{fig:genome,fig:amp,fig:montage,fig:trail}.
Because seven texts alone are too weak to support a statistical claim, we treat these numbers
as only being illustrative; \cref{sec:validation} provides corresponding measurements
over a larger held-out corpus.

\begin{table}[tb]
\caption{Structural amplification across the case-study corpus, grouped by
category. ``Bytes'' is the total number of bytes in the text; ``amp./char'' is the
number of motion regions the body traverses per source character, averaged over
40 generated runs of each text; ``entropy'' is calculated over the region
distribution. Because the engine reaches almost every requested region by direct
motion matching, the regions it traverses are ordinarily just the bytes of the text,
and the body performs the encoding itself; the occasional inserted bridge is what
lifts these averages slightly above the byte count.}
\label{tab:cases}
\centering
\footnotesize
\setlength{\tabcolsep}{4pt}
\begin{tabular}{@{}lrrrr@{}}
\toprule
Text                       & Chars & Bytes & Amp./char & Entropy \\
\midrule
\multicolumn{5}{@{}l}{\itshape ASCII English, canonical and marginalized}\\
\quad Sonnet                   & 39 & 39 & 1.01 & 3.94 \\
\quad Sojourner Truth          & 16 & 16 & 1.02 & 3.45 \\
\addlinespace[2pt]
\multicolumn{5}{@{}l}{\itshape Scripts UTF-8 encodes with several bytes}\\
\quad Hindi / Devanagari       & 13 & 35 & 2.84 & 2.86 \\
\quad Cherokee (Indigenous)    & 10 & 28 & 2.81 & 3.09 \\
\addlinespace[2pt]
\multicolumn{5}{@{}l}{\itshape Machine registers}\\
\quad Source code (Python)     & 52 & 52 & 1.00 & 4.53 \\
\quad Traceback                & 70 & 70 & 1.00 & 4.49 \\
\addlinespace[2pt]
\multicolumn{5}{@{}l}{\itshape Almost no lexical content}\\
\quad SOS / Morse              & 11 & 11 & 1.00 & 1.44 \\
\bottomrule
\end{tabular}
\end{table}

\subsection{Every text has a body}
The genomes in \cref{fig:genome} are immediately and legibly different from one
another. This confirms the instrument's basic claim: structure alone, with meaning
withheld, is enough to give every text a distinct physical identity. Across the
corpus, the mean pairwise edit distance between genomes is 44.3 regions, which
quantitatively supports what the figure shows. Furthermore, because the
byte-to-region mapping is fully deterministic, the same text always yields the same
genome, so this structural identity is stable and repeatable. The engine then
realizes that genome through motion matching, selecting specific frames and, where
needed, bridging regions to keep the body continuous; the exact frames of any
single performance depend on that search and can vary between runs. In practice the
engine reaches almost every requested region directly: across 40 runs of each
of the seven texts, no bridge was required in 92\% of runs, although an
occasional run inserts several. It is therefore the genome, rather than any single
performance, that serves as the text's fingerprint. \Cref{fig:teaser} makes the point at a glance, presenting one
characteristic pose for each of the seven texts.

\subsection{Amplifying what ASCII marginalizes}
\begin{figure*}[!t]
  \centering
  \includegraphics[width=0.95\linewidth]{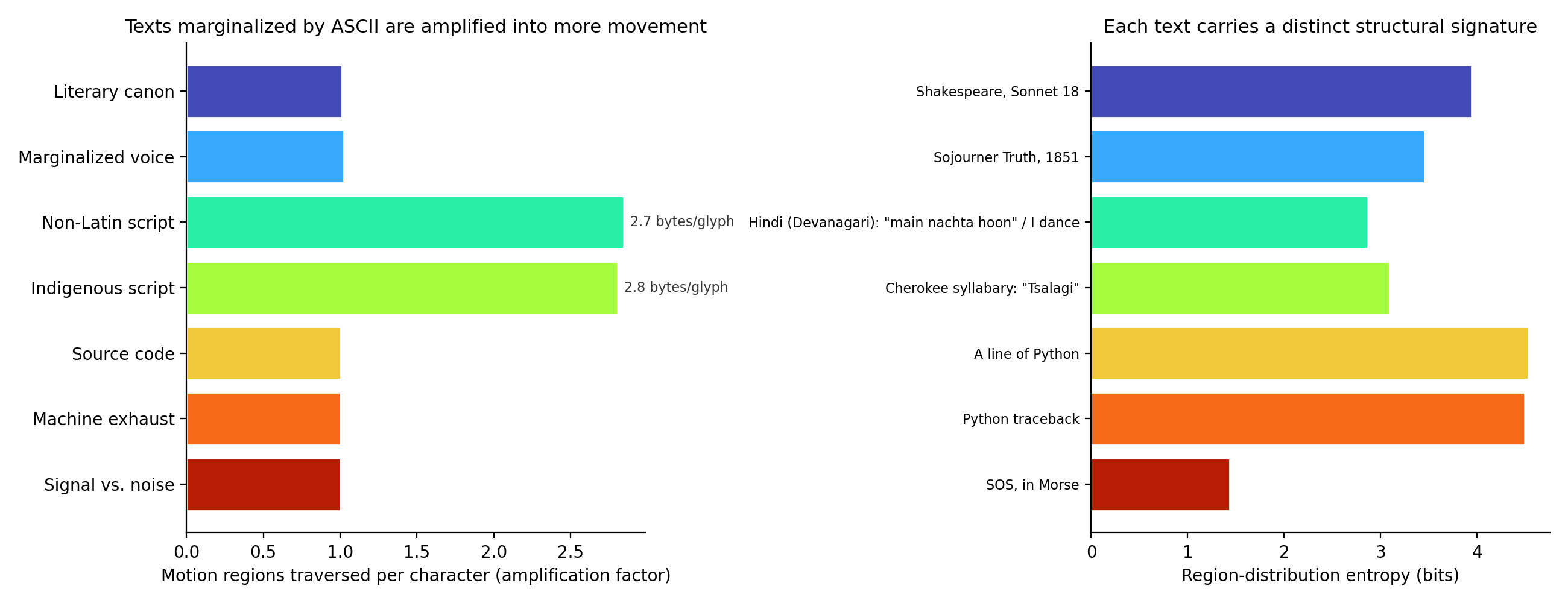}
  \caption{Left: motion regions traversed per character. Texts in scripts that
  UTF-8 encodes with several bytes per glyph are amplified into markedly more
  movement. Right: each text's region-distribution entropy, which is its structural
  signature. Source code and machine exhaust are the most varied, while the Morse
  distress call is the most repetitive.}
  \label{fig:amp}
\end{figure*}

The most pointed result is also the simplest. ASCII English spends roughly one
byte per character, so the canonical and the marginalized English texts amplify at
approximately one region per character. However, UTF-8 spends two to four bytes on
the scripts it treats as exceptions to a Latin default. The Devanagari phrase
meaning ``I dance'' (13 characters) and a Cherokee phrase (10 characters)
therefore unfold into 35 and 28 regions, which are amplification factors of around 2.8 (\cref{fig:amp}).
\Cref{fig:trail} visualizes this directly, presenting the
full generated dance for the Devanagari phrase as a chronophotographic sweep beside
those of a sonnet and a traceback; the instrument never reads that the phrase means
``I dance'', yet the byte structure alone produces a complete choreographic arc. The very encoding overhead
that makes these scripts ``expensive'' and peripheral to computing becomes, in our
instrument, a surplus of embodied expression: the overlooked script is given more
body, not less. We are careful not to overstate this result. The system does not
understand these languages, and its movement vocabulary carries its own cultural
biases, which we discuss in \cref{sec:discussion}. However, the inversion is real
and demonstrable, and we find it a productive provocation, a visualization in which
the cost of marginalization is re-read as amplitude.

\subsection{The dance of a sonnet versus an error log}
\label{sec:contrast}
\begin{figure*}[!t]
  \centering
  \includegraphics[width=0.95\linewidth]{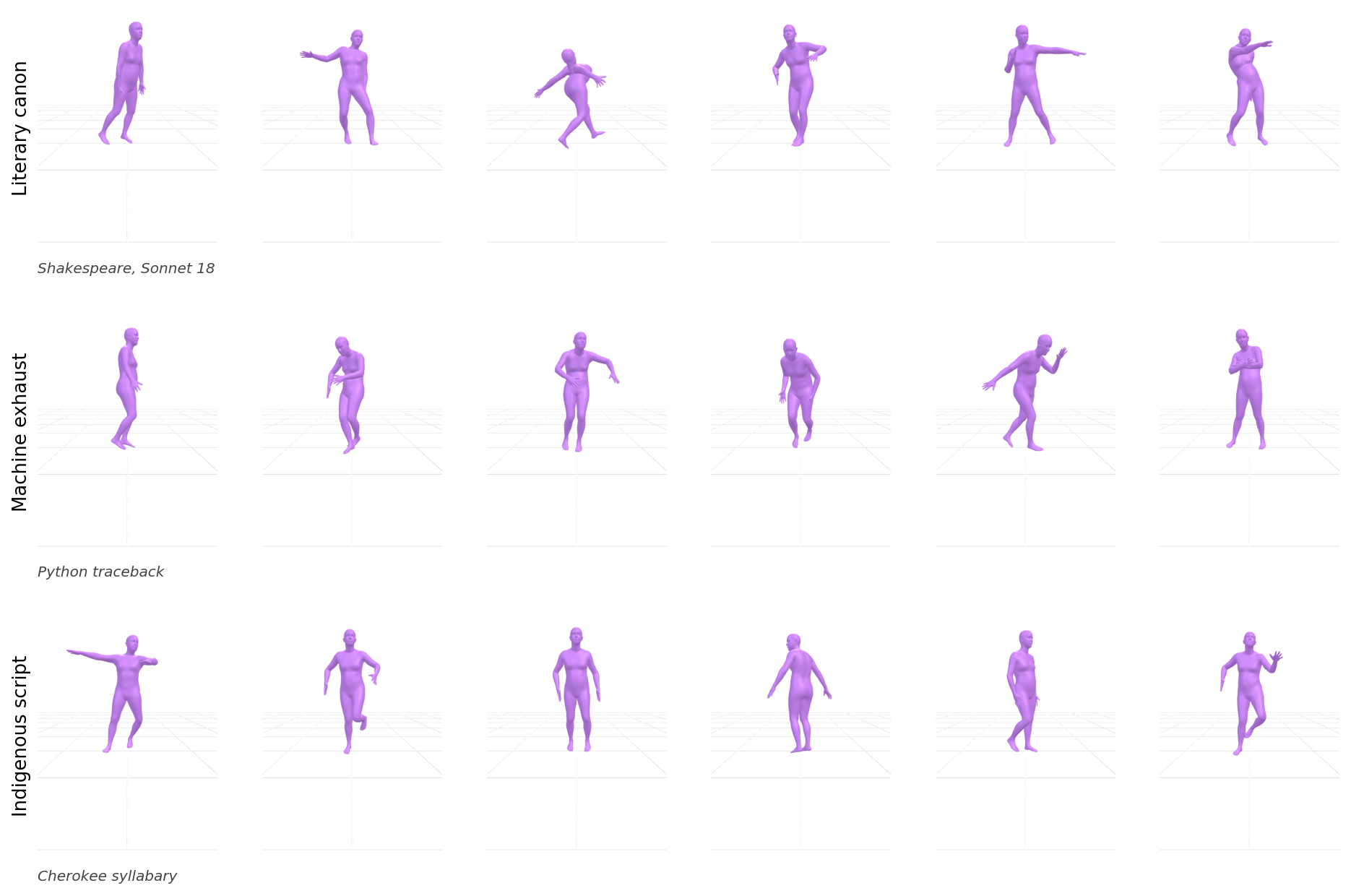}
  \caption{Physicalized choreography: time-sampled keyframes from the same
  engine driven by three different texts. The sonnet (top) is comparatively legato,
  the machine traceback (middle) is percussive and high in entropy, and the
  Cherokee syllabary (bottom) moves through long, dense phrases. These are renders
  of the actual generated motion, not schematic illustrations.}
  \label{fig:montage}
\end{figure*}

\Cref{fig:montage} renders three texts as time-sampled keyframes from the same
engine, and the contrast is borne out in the body. The sonnet, built from
lowercase Latin letters that cluster in a narrow byte range, moves in relatively
legato and recurring postures. The Python traceback, which mixes punctuation,
digits, mixed case, and the bracketed debris of a crash, carries one of the
highest structural entropies in the corpus (4.49 bits) and reads as jagged and
percussive, a body stuttering through exception handling. The Cherokee syllabary,
composed entirely of high continuation bytes, sweeps through the high regions of the
codebook in long and dense phrases. Thus, the same instrument, given
three different structures, choreographs three visibly different bodies.

\subsection{What counts as signal}
Our corpus also probes the question of what a signal actually means. ``SOS'' in Morse is a
message that is nothing but a call to be heard, and it holds almost no lexical
content. Yet the instrument gives it the lowest-entropy genome in the set (1.44
bits): a short, insistent, repeating pulse (\cref{fig:genome}, bottom row). A
semantic model would find nothing to say about three dots, three dashes, and three
dots. Our structural amplifier instead produces a clear rhythmic signature. This
raises the question of which of the two systems has actually registered the signal.

\section{Does the Instrument Work?}
\label{sec:validation}

Amplification is only meaningful if the instrument is faithful and the body it
drives is convincing. We therefore report a compact technical validation, and we
keep it deliberately brief. Except where noted, it is measured over a held-out
corpus of 570 lines of song lyrics, poetry, quotations, and jokes.
First, on the novelty of generated dance sequences: the sequences generated from text
contain choreographic phrases that are almost entirely absent from the training
corpus, with $99.9\%$ of three-gram and $100.0\%$ of four-gram region transitions
novel relative to AIST++. This confirms that the instrument composes rather than
retrieves. Second, on distinctness: genomes generated from different inputs diverge
sharply, with a mean region-sequence edit distance of $30.6$ over the 4950 pairs
formed by 100 randomly drawn held-out lines, rising to $34.0$ over the corpus as a
whole. Note that we define edit distance as the number
of single-region insertions, deletions or substitutions needed to turn one genome
into another. Additionally, text-driven motion carries a markedly higher mean
kinetic energy than the
training data, of $0.37$ versus $0.07$. This demonstrates that amplifying a text's
structure forces the body through dynamic changes of tempo. Third, on
traversability: the plausibility graph connects $54\%$ of all region pairs
directly, and Dijkstra bridging keeps detours short, with a mean of $0.56$ inserted
bridges per transition and most transitions requiring none. Finally, on fluidity:
the physics layer reduces foot-skating, the tendency of a planted foot to slide,
from $0.121$ to $0.010$, and jerk, the rate at which acceleration changes, from
$1.2\times10^{-2}$ to $8.9\times10^{-4}$, the latter below the AIST++ baseline
itself, with negligible floor penetration. The instrument is therefore both
inventive and physically credible. However, none of these numbers is offered as a
claim to state-of-the-art motion synthesis; they establish only that the amplifier
is faithful to its input and that the body it drives is one a viewer can find
plausible.

\subsection{One genome, two vocabularies}
\label{sec:vocab}
The operator is fixed by the encoding, but the regions themselves are learned.
This learned vocabulary carries the cultural specificity discussed in
\cref{sec:discussion}. To isolate this effect, we built a second instrument
differing in exactly one respect: its vocabulary is quantized exclusively from the
ballet-jazz subset of AIST++, which is roughly a tenth of AIST++. It features
its own principal components, 256 clusters, and plausibility graph. The
genome is identical in both instruments; only the performing body changes.

\begin{figure}[tb]
  \centering
  \includegraphics[width=\linewidth]{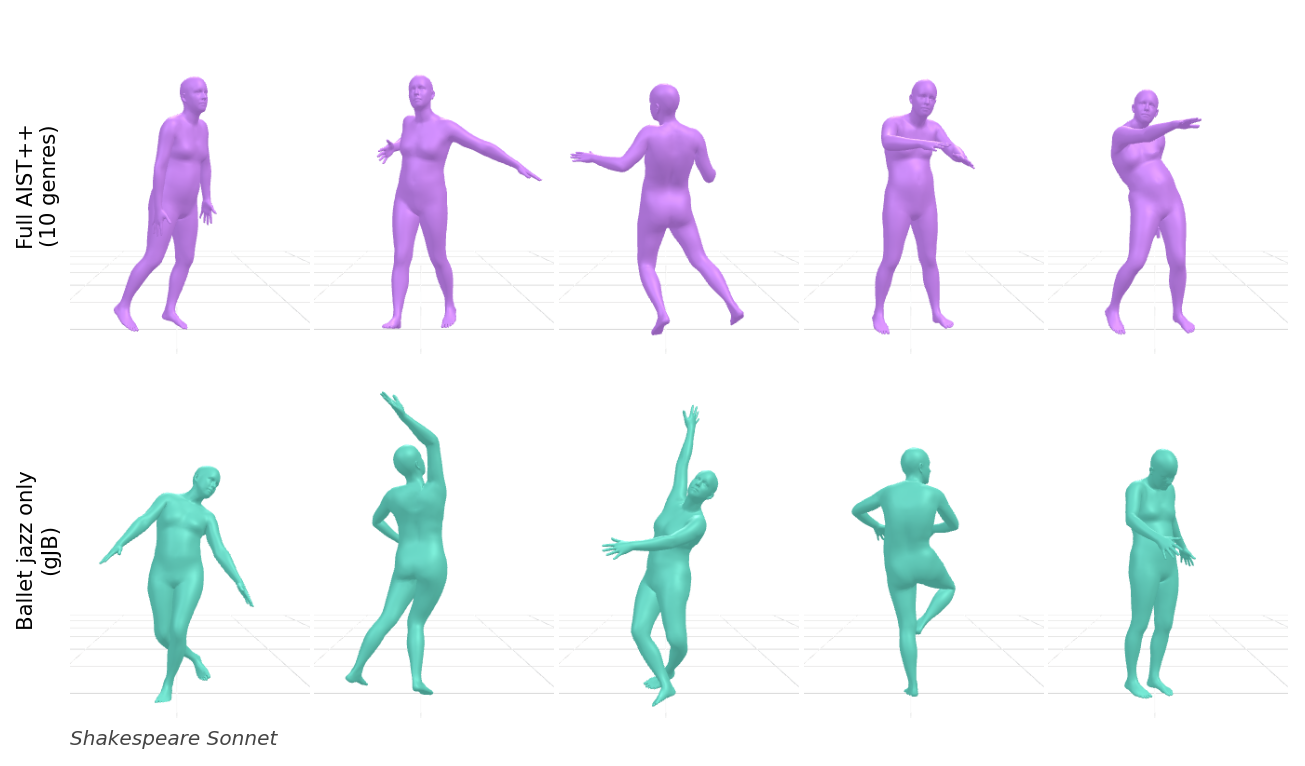}
  \caption{One genome, two bodies. The same Shakespeare sonnet, performed by
  the full AIST++ instrument (top) and an otherwise identical instrument restricted
  to the ballet-jazz subset (bottom). The numerical genome remains identical in both;
  only the manner of performance changes.}
  \label{fig:vocab}
\end{figure}

\Cref{fig:vocab} illustrates this outcome for one text. Across the case
studies, the ballet-jazz instrument moves with roughly three times the kinetic
energy of the full-corpus baseline (0.36 against 0.12) and holds its limbs 14\%
further from the pelvis (27.9 against 24.6). This aligns naturally with a
dance form built on extension and line. Because it is learned from a tenth
of the movement data, its plausibility graph is also a third as dense (0.18 against
0.54), requiring it to bridge more often. We view this as a clear division
of labor: the operator determines the score, while the vocabulary determines the
performance. A different training corpus does not yield a necessarily better instrument, but simply the same instrument speaking in a different voice. 

\section{Discussion}
\label{sec:discussion}

The central contribution of this work is less the pipeline than the inversion it
embodies. Semantic generation amplifies the part of language that is already loud,
its sanctioned meaning, and in doing so it reproduces the biases of whoever
assembled the training data~\cite{crawford2021atlas,dignazio2020datafeminism}. By
amplifying structure instead, our instrument turns up a signal that is present in
every text but attended to in almost none, and it does so on equal terms for a
sonnet and a stack trace. In this sense, the moving body becomes an instrument that
makes the quiet material of data legible.

\subsection{What generalizes, and what does not}

The operator is fundamentally indifferent to text. \Cref{eq:amp} requires only a
stream of discrete symbols and a matching vocabulary, meaning the instrument could
equally read packet captures, MIDI files, or image bytes. The display also need
not be a body; any perceptible quantized vocabulary, such as musical timbres or
lighting states, could replace our 256 motion regions. What does not generalize
is the justification for doing so. This approach is only warranted when the
dominant interpretation of a data stream discards consequential structure, much
like how standard parsers ignore UTF-8 byte economy for non-Latin scripts.
If the discarded structure is genuinely arbitrary, amplifying it produces mere
novelty rather than meaning. Distinguishing between these cases remains a
human judgment rather than a purely technical one.

\subsection{Amplification is never neutral}

At the same time, amplification is never neutral, and we resist a triumphant
reading of our own results. The movement vocabulary is learned from AIST++, a
corpus of largely studio street dance, so the body that speaks every text is
culturally specific, and the equal treatment we claim holds only with respect to
that vocabulary; \cref{sec:vocab} illustrates how much of the final result that vocabulary
is actually responsible for. Furthermore, the amplification of multibyte scripts is a property
of UTF-8, not evidence of any understanding of the languages it encodes; the
instrument can give Cherokee more movement without knowing a single word of
Cherokee. We consider this honesty to be part of the work, because an amplifier
that concealed its own coloration would be exactly the kind of system this project
sets out to question.

\subsection{Playing the instrument}

Because the core of the instrument is classical and interpretable, an artist can
actually read it. The genome is legible, the bridges are logged, and the same text
reliably yields the same genome, which the engine can replay exactly when its
random seed is fixed. This makes the system a partner that an artist can
play and argue with, rather than an oracle that must be trusted, which is closer to
an instrument than to an autocomplete. We imagine performers ``playing'' texts in
real time, curators amplifying archival documents into embodied form, and
communities choreographing words in scripts that mainstream models still render as
empty boxes.

We have also shown generated sequences to dancers informally, reporting their
responses as anecdote rather than formal evaluation. One dancer was
``surprised by how believably human the moves were'' and noted that comparing
dances from different texts made it ``clear they're expressing different
things''. They also identified specific styles, such as hip hop and
breakdancing, that were ``mixed creatively with more abstract moves''.

Finally, the natural home of this work is spatial and temporal. Beyond the page,
the instrument supports a gallery installation in which audience-supplied texts are
amplified into projected choreography in real time, as well as a multi-screen piece
that sets the dances of canonical and overlooked texts side by side. Supplementary
video renders accompany this submission.

\subsection{Future work}

Two extensions follow directly from this work. First, the vocabulary should be
learned from broader motion corpora. AIST++ consists predominantly of street dance,
and the generated body inherits this specificity. Though currently constrained by
compute and data availability, \cref{sec:vocab} shows how substantially a different
dataset would alter the results. Second, the score should be given to human
performers. Because our figures are solely machine renders, we have characterized
an instrument rather than proven it produces choreography a human would actually
want to perform. Using the genome as notation for trained dancers and comparing
their interpretation with the engine's is the natural next step for evaluation.

\section{Conclusion}
In this work, we presented an embodied visualization instrument that amplifies the
hidden structure of text into movement. By mapping raw bytes onto a quantized
vocabulary of gesture and refusing to interpret meaning, the system turns the
dancing body into a display for the quiet, material signal that semantics-first
generation discards. Across a corpus of canonical, marginalized, machine, and
Indigenous texts, it renders each input as a distinct choreographic genome, and it
amplifies the scripts that our infrastructure overlooks into more, rather than
less, embodied expression. Ultimately, we offer this instrument less as a better
way to make dances than as a way to ask, of any data stream, what we have chosen
to hear, and what we might choose to amplify instead.

\section{AI Incorporation and Usage}
The artwork is a generative system by design; the choreography in every figure is produced algorithmically by the pipeline described in \cref{sec:system}. The authors curated the inputs and the framing rather than
individual keyframes. The code was mostly written by a combination of GitHub Copilot and Claude Code, but prompts and ideas came from humans. We deliberately use classical, interpretable machine
learning, specifically PCA, K-Means, and graph search, rather than a large neural
generator, for sake of interpretability.

AI assistance was utilized for initial drafting purposes, most notably for literature review and result analysis. A combination of Gemini and Claude Code were used. All AI drafted content has been reviewed and edited by the authors, and the authors take full responsibility for the writing.

\section*{Supplemental Materials}
\label{sec:supplemental_materials}
The source code, the motion codebook and plausibility graph, the case-study
corpus, and supplementary video renders of the choreography are available at
\url{https://github.com/ML72/Choreographic-Genome}, with a visual demo at
\url{https://ml72.github.io/Choreographic-Genome/}. All figures in this paper were
generated by the system described here, and no figure is reproduced from another
source.

\acknowledgments{
We gratefully acknowledge the instructors of the Art and Machine Learning (10-615) course at Carnegie Mellon University in Spring 2026, whose insightful lectures sparked the initial inspiration for this project. We also thank the creators of the AIST++ dataset for making this work possible.}

\bibliographystyle{abbrv-doi}
\bibliography{references}

\end{document}